\documentclass[a4paper,conference]{IEEEtran}
\usepackage[dvipdfm]{graphicx}
\usepackage[dvipdfm]{color}
\begin{document}

\title{Money Distributions in Chaotic Economies}

\author{
\authorblockN{Carmen Pellicer-Lostao}
\authorblockA{Centro Polit\'{e}cnico Superior \\
Universidad de Zaragoza \\ E-50015 Zaragoza, Spain\\
Email: carmen.pellicer@unizar.es}
\and
\authorblockN{Ricardo L\'opez-Ruiz}
\authorblockA{Facultad de Ciencias \\
Universidad de Zaragoza \\ E-50009 Zaragoza, Spain\\
Email: rilopez@unizar.es}}

\maketitle

\begin{abstract}

This paper considers the ideal gas-like model of trading markets, where each individual is identified as a gas
molecule that interacts with others trading in elastic or money-conservative collisions. Traditionally this
model introduces different rules of random selection and exchange between pair agents. Real economic
transactions are complex but obviously non-random. Consequently, unlike this traditional model, this work
implements chaotic elements in the evolution of an economic system. In particular, we use a chaotic signal that
breaks the natural pairing symmetry $(i,j)\Leftrightarrow(j,i)$ of a random gas-like model. As a result of that,
it is found that a chaotic market like this can reproduce the referenced wealth distributions observed in real
economies (the Gamma, Exponential and Pareto distributions).

\end{abstract}

\IEEEpeerreviewmaketitle

\section{Introduction}
Econophysics is a relatively new discipline ~\cite{mantegna} that applies many-body techniques developed in
statistical mechanics to the understanding of self-organizing economic systems ~\cite{yako2009}. The techniques
used in this field ~\cite{yako2000}, ~\cite{chakra2000}, ~\cite{bouchaud} have to do with agent-based models and
simulations. The statistical distributions of money, wealth and income are obtained on a community of agents
under some rules of trade and after an asymptotically high number of interactions between the agents.

The conjecture of a kinetic theory with (ideal) gas-like behaviour for trading markets was first discussed in
1995 ~\cite{chakra1995} by econophysicists. This model considers a closed economic community of individuals
where each agent is identified as a gas molecule that interacts randomly with others, trading in elastic or
money-conservative collisions. Randomness is an essential ingredient in this model, where agents interact in
pairs chosen at random by exchanging a random quantity of money. The interesting point of this model is that, by
analog with Energy, the equilibrium probability distribution of money follows the exponential Boltzmann-Gibbs
law for a wide variety of trading rules ~\cite{yako2009}.

This result is coherent with real economic data up to some extent. Nowadays it is well established that the
income and wealth distributions in many countries follow a characteristic pattern. They seem to present two
phases with different richness distributions. One phase presents an exponential (Boltzmann-Gibbs) distribution
and covers about $90-95\%$ of individuals, those with low and medium incomes. The other phase, which is
integrated by the rest of the individuals with high incomes, shows a power law (Pareto). This is so in different
countries, irrespective of their differences in cultural or social structures , ~\cite{yako2001}, ~\cite{sinha},
from older societies ~\cite{abul2002} ~\cite{pareto} ~\cite{champer53} ~\cite{champer99} to modern ones
~\cite{chatter2005}, ~\cite{chakra2006}, ~\cite{yako2001b}.

This paper introduces chaotic-driven dynamics in the economic gas-like model. The objective is obtaining money
distributions similar to those observed in real economies. This is a novel approach, where the rules of
selection of agents and money transfers are no longer random, but driven by nonlinear maps in chaotic regime.
Two different scenarios are considered and the money distributions, obtained that way, are compared to the
distributions shown in real economies.

The contents of this paper are organized as follows: section 2 describes the simulation scenario. Section 3 show
the results obtained. Final conclusions are in Section 4.

\section{Scenario of Chaotic Simulation}

This paper introduces chaos in the dynamics of the economic gas-like model. This is done upon two facts
that seem particularly relevant in this purpose.

The first one is, that real Economy is not purely random. Economic transactions are driven by some specific
interest (of profit) between the interacting parts. Thus, on one hand, there is some evidence of markets being
not purely random. On the other hand, everyday life shows us the unpredictable component of real economy with
its recurrent crisis. Hence, it can be sustained that the short-time dynamics of economic systems evolves under
deterministic forces though, in the long term, these kind of systems show an inherent instability. Therefore,
the prediction of the future situation of an economic system resembles some-how to the weather prediction. It
can be concluded that determinism and unpredictability, the two essential components of chaotic systems, take
part in the evolution of Economy and Financial Markets.

The second fact is, that the transition from the Boltzmann-Gibbs to the Pareto distribution can require the
introduction of some kind of inhomogeneity that breaks the random indistinguishability among the individuals in
the market, although this is not a necessary condition in order to have such kind of transition \cite{gonzalez2009}.
Nonlinear maps can be an ideal candidate to obtain that, as they can offer quite flexible models of
evolution in the market. In particular, nonlinear maps can be easily tuned to break the pairing symmetry of
agents $(i,j)\Leftrightarrow(j,i)$, characteristic of the random model.

The simulation scenario presented here follows a traditional gas-like model. A community of $N$ agents is given
with an initial equal quantity of money, $m_0$, for each agent. The total amount of money, $M=N*m_0$, is
conserved in time. For each transaction, a pair of agents $(i,j)$ is selected, and an amount of money $\Delta m$
is transferred from one to the other. In this work two simple and well known rules are used. Both consider a
variable $\upsilon$ in the interval $(0,1)$, not necessarily random, in the following way:

\begin{itemize}
    \item \textbf{Rule 1}: the agents undergo an exchange of money, in a way that agent $i$ ends up with a
    $\upsilon$-dependent portion of the total of two agents money, ($\upsilon*(m_i+m_j)$),
    and agent $j$ takes the rest ($(1-\upsilon)*(m_i+m_j)$) ~\cite{patriarca}.
    \item \textbf{Rule 2}: an $\upsilon$-dependent portion of the average amount of the two agents money,
    $\Delta m=\upsilon*(m_i+m_j)/2$, is taken from $i$ and given to $j$ ~\cite{yako2000}. If $i$ doesn't have enough money, the transfer doesn't take place.
\end{itemize}

As there are two different simulation parameters involved in these gas-like models (the parameter for selecting
the agents involved in the exchanges and the parameter defining the economic transactions), two different
scenarios are obtained depending on the random-like or chaotic election of these parameters. These scenarios are
the following:

\begin{itemize}
\item \textbf{Scenario I}: random selection of agents with chaotic money exchanges. \item \textbf{Scenario II}:
chaotic selection of agents with random money exchanges.
\end{itemize}

To produce chaotically a pair of agents $(i,j)$ for each interaction, a 2D nonlinear map under chaotic regime is
considered. The pair $(i,j)$ is easily obtained from the coordinates of a chaotic point at instant $t$,
$X_t=[x_t, y_t]$, by a simple float to integer conversion ($x_t$ and $y_t$ to $i$ and $j$, respectively).
Additionally, to obtain a chaotic money exchange, a float number $\upsilon$ in the interval $[0,1]$ is obtained
form the coordinates of a chaotic point at instant $t$ by taking one or a combination of both. This number
produces the chaotic quantity of money $\Delta m$ that is traded between agents $i$ and $j$.

The simulation scenario uses two particular 2D nonlinear maps under chaotic regime. These are the Henon map
described in ~\cite{henon} (\ref{equ_1}) and the Logistic bimap described in ~\cite{ric} as model (a)
(\ref{equ_2}) . These maps are given by the following equations:

\begin{equation}
\begin{array}{l}
T_H:\Re\times\Re\longrightarrow\Re\times\Re
\\ x_{t+1} = ax_t^2 + y_t + 1,
\\ y_{t+1} = bx_t.
\end{array}
\label{equ_1}
\end{equation}

\begin{equation}
\begin{array}{l}
T_A:[0,1]\times[0,1]\longrightarrow[0,1]\times[0,1]
\\ x_{t+1} =\lambda(3y_t+1)x_t(1-x_t),
\\ y_{t+1}=\lambda(3x_t+1)y_t(1-y_t).
\end{array}
\label{equ_2}
\end{equation}

These maps show chaotic behaviour for some values of their parameters. In this work the following values are
considered: the H\'enon map in its canonical with $a=1.4$ and $b=0.3$ and the Logistic bimap when $\lambda_a$
and $\lambda_b$ are in the interval $[1.032, 1.0843]$, where chaotic regime can be obtained.

To conclude the description of the simulation scenario, it is interesting to point out here that the two rules
of trade considered here exhibit different symmetry of exchange. Rule $2$ is asymmetric for it takes money from
$i$ agent and gives it to the other. When considering the chaotic maps, the Henon map is asymmetric and the
Logistic bimap exhibits symmetric behaviour along the diagonal axis (straight line $y=x$) when
$\lambda_a$=$\lambda_b$. This symmetry is modified in the simulations to show how asymmetry can affect the final
money distributions.

\section{Asymptotic Money Distributions}

The (ideal) gas-like model, as described in the precedent sections shows an equilibrium distribution of money
that follows the exponential Boltzmann-Gibbs law for a wide variety of trading rules ~\cite{yako2009,lopezruiz2008}.

In this section, chaos is introduced in this model. Different scenarios are presented depending on the way this
chaoticity is established. These scenarios where first proposed by the authors in ~\cite{iccs2009}. The obtained
asymptotic money distributions are presented, compared to the exponential and discussed in order to provide
illuminating ideas about the cause of these particular results.

\subsection{Scenario I: random agents/chaotic trade}
In this scenario, the selection of agents is random as in the traditional gas-like model. Individuals don't have
preferences in choosing an interaction partner, commercial transactions are nor restricted, nor driven by
particular interests. But unlike this model, the exchange of money is forced to evolve according to nonlinear
patterns. Economically, this means that the exchange of money has a deterministic component, although it varies
chaotically. Put it in another way, the prices of products and services are discrete and evolve in a limited
range, in a complex way.

A community of $N=500$ individuals is considered with an initial quantity of money of $m_0=1000\$$. This
community takes a total time of $400000$ transactions. For each transaction two random numbers from a standard
random generator are used to select a pair of agents. Additionally, a chaotic float number is produced to obtain
the float number $\upsilon$ in the interval $[0,1]$. The value of $\upsilon$ is calculated as $|x_i|/1.5$ for
the H\'enon map and as $x_i$ for the Logistic bimap. This value and the rule selected for the exchange determine
the amount of money $\Delta m$ that is transferred from one agent to the other.

\begin{figure}
\centerline{\includegraphics[width=8cm]{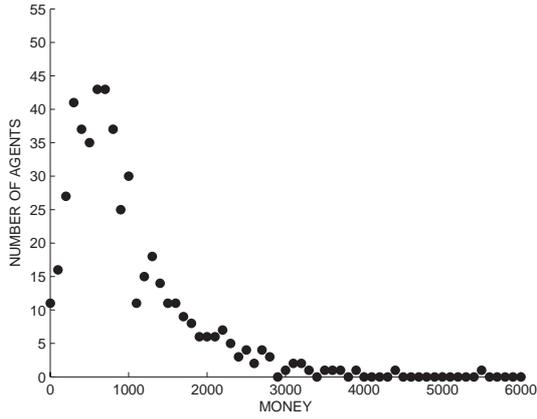}} \caption{Chaotic trade selection with Logistic
Bimap and Rule 1} \label{fig1}
\end{figure}

\begin{figure}
\centerline{\includegraphics[width=8cm]{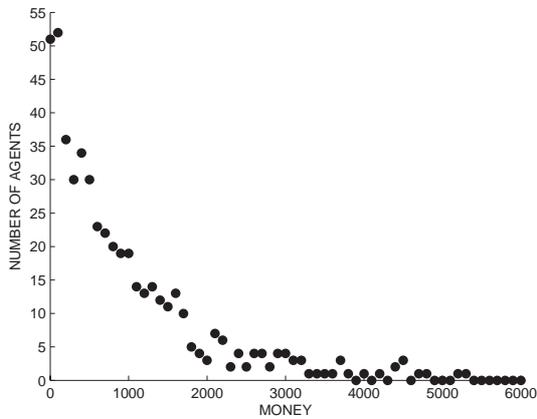}}
 \caption{Chaotic trade selection with Logistic Bimap and Rule 2} \label{fig2}
\end{figure}

Different cases are considered, taking the H\'enon chaotic map or the Logistic bimap. Rules 1 and 2 are also
considered. New features appear in this scenario. These can be observed in Fig.\ref{fig1} and Fig.\ref{fig2}.

The first feature is that the chaotic behavior of $\upsilon$ is producing a different final distribution for
each rule. Rule 2 is still displaying the exponential shape, but Rule 1 gives a different asymptotic function
distribution, it becomes a Gamma-like distribution. It presents a very low proportion of the population
in the state of poorness, and a high percentage of it in the middle of the richness scale, near to the value of
the mean wealth. Rule 1 seems to lead to a more equitable distribution of wealth.

Basically, this is due to the fact that Rule 2 is asymmetric. Each transaction of Rule 2 represents an agent $i$
trying to buy a product to agent $j$ and consequently agent $i$ always ends with the same o less money. On the
contrary, Rule 1 is symmetric and in each interaction both agents $(i,j)$, as in a joint venture, end up with a
division of their total wealth. Now, think in the following situation: with a fixed $\upsilon$, let say
$\upsilon=0.5$, Rule 1 will end up with all agents having the same money as in the beginning, $m_0=1000\$$.
Using a chaotic evolution of $\upsilon$ means restricting its value to a defined region, that of the chaotic
attractor. Consequently, this is enlarging the distribution around the initial value of $1000\$$ but it does not
go to the exponential as in the random case ~\cite{patriarca}.

\subsection{Scenario II: chaotic agents/random trade}
In this scenario, the selection of agents chaotic dynamics. Individuals do have preferences in choosing
an interaction partner, driven by their particular interests. On the other hand the exchange of money is random,
meaning that prices of products and services are not limited and distributed uniformly in the market.

A community of $N=500$ agents with initial money of $m_0=1000\$$ is taken and the chaotic map variables $x_t$
and $y_t$ will be used as simulation parameters. This community takes a total time of $400000$ transactions. For
each transaction two chaotic floats in the interval $[0,1]$ are produced. The value of these floats are
$|x_t|/1.5$ and $|y_t|/0.4$ for the H\'enon map and $x_t$ and $y_t$ for the Logistic bimap. These values are
used to obtain $i$ and $j$ as described in previous section. Additionally, a random number from a standard
random generator are used to obtain the float number $\upsilon$ in the interval $[0,1]$. The value of $\upsilon$
and the selected rule determine the amount of money $\Delta m$ that is transferred from one agent to the other.

\begin{figure} [h]
\centerline{\includegraphics[width=8cm]{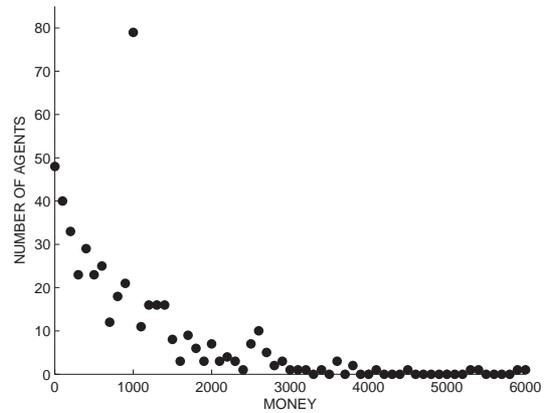}}
 \caption{Chaotic agents selection with Logistic bimap and Rule 1} \label{fig3}
\end{figure}

\begin{figure} [h]
\centerline{\includegraphics[width=8cm]{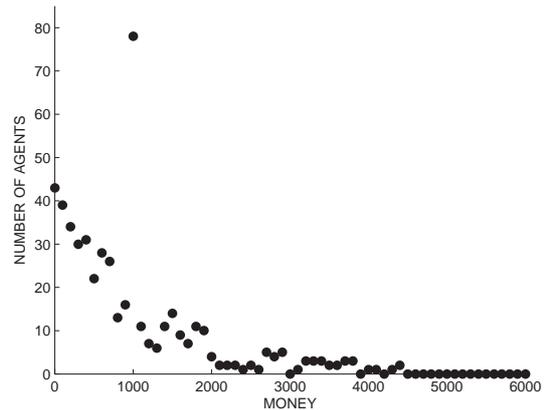}}
 \caption{Chaotic agents selection with symmetric Logistic bimap and Rule 2} \label{fig4}
\end{figure}

Different cases are considered, taking the H\'enon chaotic map or the Logistic bimap. Rules 1 and 2 are also
considered. The results can be observed in Fig.\ref{fig3} and Fig.\ref{fig4}.

As a result, an interesting point appears in this scenario with both rules. This is the high number of agents
that keep their initial money. The reason is that they don't exchange money at all. The chaotic numbers used to
choose the interacting agents are forcing trades between a deterministic group of them and hence some commercial
relations result restricted.

In can be observed in Fig.\ref{fig3} and Fig.\ref{fig4}, that the asymptotic distributions in this scenario
again resemble the exponential function. When Logistic bimap is symmetric, it produces the effect of behaving as
the gas-like model but with a restricted number of agents.

Amazingly, when the Logistic bimap becomes asymmetric and Rule 2 is used, it leads to a distribution with a
heavy tail, a Pareto-like distribution. This can be clearly seen if the value of the parameter $\lambda$ is
varied as described in Table I. In this simulation Cases the community of agents is incremented for higher
resolution to $N=5000$ individuals. Consequently the total time of transactions is also increased to
$5\times10^7$ to guarantee an asymptotic situation.

\begin{table} [h]
\label{table1}
\begin{center}
\begin{tabular}{|c|c|c|c|c|}
   \hline
  CASE & $1$ & $2$ & $3$ & $4$\\
  \hline
  $\lambda_a$ & $1.032$ & $1.032$ & $1.032$ & $1.032$\\
  \hline
  $\lambda_b$ & $1.032$ & $1.03781$ & $1.04362$ & $1.049430$\\
  \hline
\end{tabular}
\end{center}
\end{table}

\begin{table} [h]
\label{table2}
\begin{center}
\begin{tabular}{|c|c|c|c|c|}
   \hline
  CASE & $5$ & $6$ & $7$ & $8$\\
  \hline
  $\lambda_a$ & $1.032$ & $1.032$ & $1.032$ & $1.032$\\
  \hline
  $\lambda_b$ & $1.06105$ & $1.07267$ & $1.07848$ & $1.08429$\\
  \hline
\end{tabular}
\end{center}
\caption{Different cases considered in the Logistic bimap to modify symmetry in the selection of agents.}
\end{table}

\begin{figure} [h]
\centerline{\includegraphics[width=8cm]{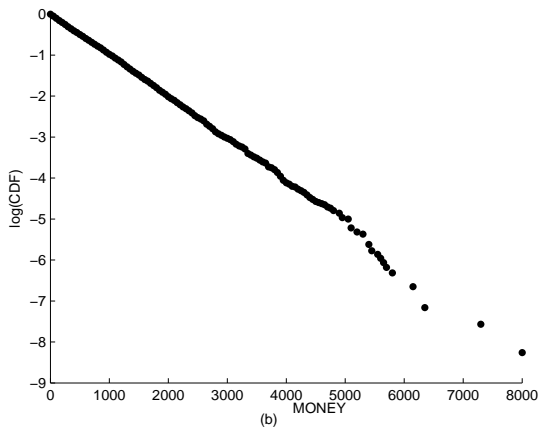}} \caption{Cumulative distribution function (CDF) of CASE 1 in Table I is drawn for the community of real participants ($3867$ agents).} \label{fig5}
\end{figure}

The resulting money distributions are then obtained as the $\lambda_b$ varies from $1.032$ to $1.08429$. When
the passive agents are removed of the model, one can obtain the money distribution of the interacting agents.
Fig. \ref{fig5} shows the cumulative distribution function (CDF) obtained for the symmetric case. Here the
probability of having a quantity of money bigger or equal to the variable MONEY, is depicted in natural log
plot, showing clearly the exponential distribution.

When $\lambda_b$ varies from $1.032$ to $1.08429$ it is observed that the number of non-participants decreases. This is because the chaotic map expands and its resulting projections on axis $x$ and $y$ grow in range, taking a greater group of $i$ and $j$ values is computed. Taking these non-participants off the final money distributions, and so their money too, one can obtain the final CDF's for the different values of $\lambda_b$.
When these distributions are depicted an interesting progression is shown. As $\lambda_b$ increases, these
distributions diverge from the exponential shape.

\begin{figure} [h]
\centerline{\includegraphics[width=8cm]{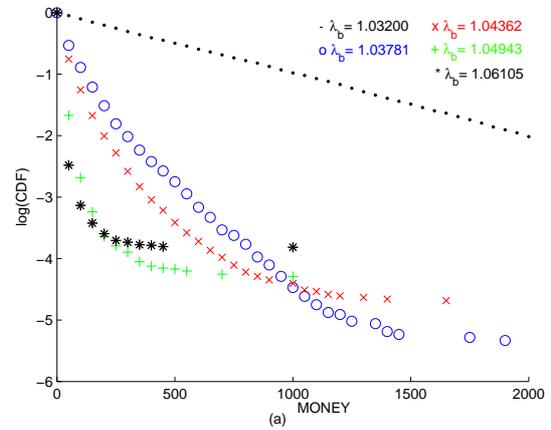}} \caption{Representations up to $2000\$$ dollars of the final
CDF's obtained for simulation cases 1,2,3,4 and 5 (see Table I).} \label{fig6}
\end{figure}

Fig. \ref{fig6} shows the representation of simulation cases $1$ to $5$ in a natural log plot up to a range of
$\$2000$. It can be appreciated that as $\lambda_b$ increases, the straight shape obtained for the symmetric
case bends progressively, the probability of finding an agent in the state of poorness increases. It also can be
seen that for cases 3 and 4, no agent can be found in a middle range of wealth (from $1000\$$ to $2000\$$). This
means that the distribution of money is becoming progressively more unequal.

\begin{figure} [h]
\centerline{\includegraphics[width=8cm]{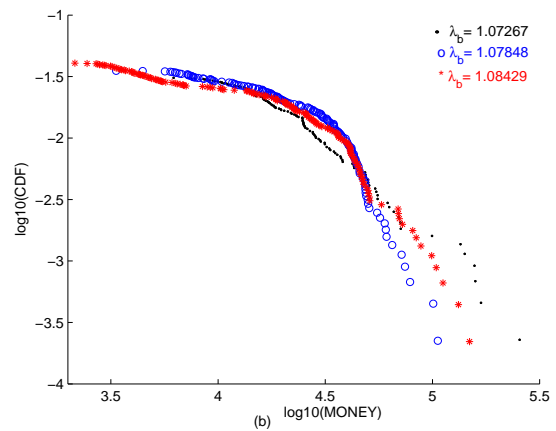}} \caption{Representations from $2000\$$ dollars of the final
CDF's obtained for simulation cases 6,7, and 8 (see Table I).} \label{fig7}
\end{figure}

In Fig. \ref{fig7} the CDF's for simulation cases $6$ to $8$ are depicted from a range of $2000\$$ and in double
decimal logarithm plot. Here, a minority of agents reach very high fortunes, what explains, how other majority
of agents becomes to the state of poorness. The data seem to follow a straight line arrangement for case $6$,
which resembles a Pareto distribution. Cases $7$ and $8$ show two straight line arrangements which can also be
adjusted to two Pareto distributions of different slopes.

Fig. \ref{fig6} and \ref{fig7} show how the community is becoming more unequal as $\lambda_b$ increases. The
rich gets richer as the asymmetry of the chaotic selection of agents increases and the final amount of money of
this class is almost the total money in the system. In the other side of the society, the proportion of the
population that ends in the state of poverty increases as as $\lambda_b$ increases.

What is happening here is, that the asymmetry of the chaotic map is selecting a set of agents preferably as
winners for each transaction ($j$ agents). While others, with less chaotic luck become preferably looser ($i$
agents). This is due to the asymmetry of Rule 2, where agent $i$ always decrements his money, and the asymmetry
of coordinates $x_i$ and $y_i$ in the Logistic bimap used for the selection of agents. This double asymmetry
makes some agents prone to loose in the majority of the transactions, while a few others always win.

\section{Conclusions}
This work introduces chaotic dynamics in economic (ideal) gas-like models for trading markets. It shows that
introducing chaotic parameters in two different scenarios leads to the referenced money distributions seen in
real economies (the Gamma, Exponential and Pareto distributions).

New and interesting results are observed in these scenarios, in the sense that restriction of commercial
relations is observed, as well as a different asymptotic wealth distribution depending on the rule of money
exchange. It seems that the asymmetry of the trading rule and also of the chaotic selection of agents leads to
less equitable distributions of money, meaning that this type of the markets operate under ''unfair`` or
asymmetric conditions.

More over, under these assumptions, it illustrates how a small group of people can be chaotically destined to be
very rich, while the bulk of the population ends up in state of poverty. This may resemble some realistic
conditions, showing how some individuals can accumulate big fortunes in trading markets, as a natural
consequence of the intrinsic asymmetric conditions of real economy.

{\bf Acknowledgements} The authors acknowledge some financial support by Spanish grant
DGICYT-FIS200612781-C02-01.

\end{document}